\documentclass[twocolumn,showpacs,preprintnumbers,amsmath,amssymb,aps,prl]{revtex4-1}
\usepackage{graphicx}
\usepackage{dcolumn}
\usepackage{bm}
\usepackage[shortlabels]{enumitem}
\begin{document}
.
\preprint{LA-UR-15-26631}

\title{Atomic Motion from the Mean Square Displacement in a Monatomic Liquid}
\author{Duane C. Wallace}
\affiliation{Theoretical Division, Los Alamos National Laboratory, 
Los Alamos, New Mexico 87545}
\author{Eric D. Chisolm}
\affiliation{Theoretical Division, Los Alamos National Laboratory, 
Los Alamos, New Mexico 87545\\
 and Descartes Labs, Los Alamos, NM 87544}
\author{Giulia De Lorenzi-Venneri}
\affiliation{Theoretical Division, Los Alamos National Laboratory, 
Los Alamos, New Mexico 87545}

\date{\today}
\begin{abstract}
V-T theory is constructed in the many-body Hamiltonian formulation, and differs at the foundation from current liquid dynamics theories. In V-T theory the liquid atomic motion consists of two contributions, normal mode vibrations in a single 
representative potential energy valley, and transits, which carry the system across boundaries between valleys. The mean square displacement time correlation function (the MSD) is a direct measure of the atomic motion , and our goal is to determine if the V-T formalism can produce a physically sensible account of this motion. We employ molecular dynamics (MD) data for a system representing liquid Na, and find the motion evolves in three successive time intervals: On the first ``vibrational" interval, the vibrational motion alone gives a highly accurate account of the MD data; on the second ``crossover" interval, the vibrational MSD saturates to a constant while the transit motion builds up from zero; on the third ``random walk" interval, the transit motion produces a purely diffusive random walk of the vibrational equilibrium positions. This motional evolution agrees with, and adds refinement to, the MSD atomic motion as described by current liquid dynamics theories.
\end{abstract}

\maketitle

The mean square displacement time correlation function, abbreviated MSD, is a well known liquid theoretical property, related to the self-diffusion coefficient $D$ by the Einstein random walk model \cite{Einstein}, and also by an equivalent Green-Kubo equation for $D$ (pp.184-185 of \cite{HMcD_3rded}). To evaluate $D$ for a real liquid, one first calculates a long and heavily averaged molecular dynamics (MD) segment of the MSD, denoted $X_{MD}(t)$ where $t$ is time, for example from first-principles density functional theory (DFT), or from \emph{a priori} interatomic potentials. 
The $X_{MD}(t)$ segment contains a short initial period of non-diffusive motion, then the pure diffusive motion takes over and makes $X_{MD}(t)$ linear in $t$, where theory prescribes $\dot{X}_{MD}(t)=6D$~\cite{Einstein,HMcD_3rded}. One therefore fits a straight line to $X_{MD}(t)$ vs $t$, with the initial transient period omitted from the fitting, and evaluates $D$ from the straight line slope. The MD segment allotted to the straight line fit is typically 1000 times longer than the transient period; on a graph of $X_{MD}(t)$ with its  fitted straight line, the initial transient is generally unobservable. Nevertheless, as we shall see, the initial transient provides more information about the atomic motion than does  the diffusive random walk. 

It has long been recognized that time correlation functions show two stages of response, a short time response from particle interaction events, and a long time response associated with many particle collective effects (p.\emph{viii} and Figs. 4.6-4.8 of \cite{BoonYip1980}). This picture remains but is being refined, as in mode-coupling theory, which accounts for a time correlation function as a process of decay of fluctuations into related modes of motion (Sec. 4.1 and 4.2 of  \cite{Gotze2009}; Sec. 9.5 of \cite{HMcD_3rded}). Mode coupling theory is deeply involved in studies of the glass transition \cite{Gotze2009,BinderKob2005}; those studies often include normal liquids as a limiting case, and they provide us with the following motional description of the liquid MD data for the MSD\cite{Kob1999}: Ballistic motion at short times, with $X_{MD}(t) \propto t^{2}$, and diffusive motion at long times, with $X_{MD}(t) \propto t$. This behavior is common in a broad range of liquid types, for example in binary LJ and silica systems (Fig.~3 of \cite{Kob1999}), complex hard sphere systems \cite{Tokuyama2008}, a one-component LJ system \cite{Hoang2013}, and in first-principles MD simulations for Al \cite{JaksePasturel2013}. We shall compare our present results for the atomic motion with this established description in the conclusion. 

In contrast, V-T theory follows the classic many body Hamiltonian formulation; this is an interesting option for liquid dynamics, because that formulation already encodes in a common language much of our understanding of condensed matter physics \cite{Pines_1997_MBP, Pines_1997_EES, Kittel1963,Harrison, Ashcroft_Mermin, Bloch, Glyde_book, SPCL}. We start with the tractable Hamiltonian for harmonic vibrational motion within a single many atom potential energy valley of the class of valleys the system visits in the liquid state. To account for the liquid diffusive properties, we add transit motion, where each transit is the correlated motion of a small local group of atoms that carries the system across the boundary between two potential energy valleys. Such transits have been observed in low temperature equilibrium MD trajectories \cite{Obs_of_tr_2001}. The vibrational Hamiltonian is calculated from first principle, transits are modeled in terms of adjustable parameters, and a time correlation function is calculated directly from the combined motions. This formulation differs at the foundation from current liquid dynamics theories, as exemplified by mode-coupling theory.

In this first application of V-T theory to the MSD, our goal is to determine if the many-body formulation can produce a physically sensible account of the atomic motion underlying the MSD. Our system represents
liquid Na at 395~K, in terms of a well tested interatomic potential from pseudopotential perturbation theory (Figs. 1.1, 17.3 19.1-19.3, 20.1 of \cite{SPCL}).

$X_{MD}(t)$   is calculated directly from the theoretical expression for the MSD,
\begin{equation} \label{eq1}
X_{MD}(t) = \frac{1}{N} \sum_K \left < [   {\bf r}_{K}(t) - {\bf r}_{K}(0)]^{2} \right >_{MD},
\end{equation}
where  ${\bf r}_{K}(t)$ is the position of atom $K$ for $K=1,\dots,N$,  and  $<\dots>_{MD}$ is the average over  $t=0$ start times. The V-T theory expression, $X_{VT}(t)$, has vibrational and transit terms plus an interaction, written as
\begin{equation} \label{eq2}
X_{VT}(t) = X_{vib}(t) + X_{tr}(t) + X_{int}(t). 
\end{equation}
Our premise is that all significant contributions are contained in $X_{vib}(t)$ or $X_{tr}(t)$, while $X_{int}(t)$ expresses residual errors that can be neglected or trivially approximated. Two interaction examples are discussed in this report.

\begin{figure} [h]
\includegraphics [width=0.35\textwidth]{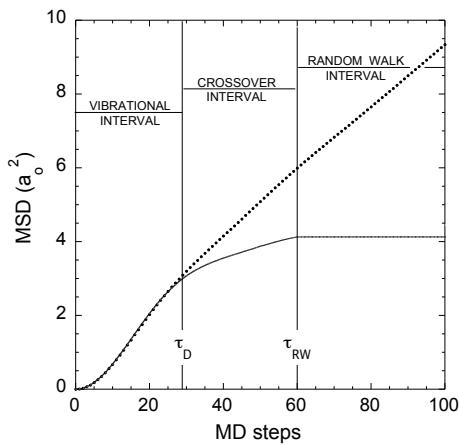}
\caption{Dots are $X_{MD}(t)$ and  line is  $X_{vib}(t)$. Each time  interval corresponds to specific underlying atomic motions; the random walk continues forever. The slope discontinuity in $X_{vib}(t)$ at $\tau_{RW}$ is explained in the text.}
\label{fig1}
\end{figure}
Fig.~\ref{fig1}  presents a comparative graph of $X_{MD}(t)$  and $X_{vib}(t)$, from which $X_{tr}(t)$ can be visualized.
The Figure is divided into three time intervals, which correspond to distinct atomic motions that underlie the MSD. The vibrational interval recognizes the remarkable property that $X_{MD}(t)$ is in near perfect agreement with $X_{vib}(t)$ up to time $\tau_{D}$ , the delay time. $\tau_{D}$ is defined as the time when $X_{MD}(t)$ moves away from $X_{vib}(t)$, hence $\tau_{D}$ is qualitative and may be chosen with some leeway. Our calibration of $\tau_{D}$ and the remaining parameters is listed in Table I.  $\tau_{RW}$ is defined as the time when $X_{MD}(t)$ reaches its ultimate straight line behavior, and will be calibrated below.

 To derive $X_{vib}(t)$ we write 
\begin{equation} \label{eq4}
{\bf r}_{K}(t) = {\bf R}_{K} + {\bf u}_{K}(t),
\end{equation}
where ${\bf R}_{K}$ is the equilibrium position and ${\bf u}_{K}(t)$ is the vibrational displacement. Following  Eq.~(\ref{eq1}) we have
\begin{equation} \label{eq5}
X_{vib}(t) =\frac{1}{N} \sum_{K} \left < [   {\bf u}_{K}(t) - {\bf u}_{K}(0)]^{2} \right >_{vib}.
\end{equation}
Now with the vibrational  equation of motion for ${\bf u}_{K}(t)$, Eq.~(\ref{eq5}) becomes  
\begin{equation} \label{eq6}
X_{vib}(t)=\frac{6 k_{B}T}{M} \frac{1}{3N-3}\sum_{\lambda} \frac{1-\cos \omega_{\lambda}t}{\omega_{\lambda}^2},
\end{equation}
where $T$ is the temperature, $M$ is the atomic mass, and $\omega_{\lambda}$ are the frequencies of the normal  modes $\lambda = 1, \dots, 3N-3$, having omitted the three modes for which $\omega_{\lambda}=0$. Eq.~(\ref{eq6}) 
is explicit and tractable; it depends on the frequency spectrum but not on the mode eigenvectors. The procedure for calculating the vibrational Hamiltonian parameters is general, and calculation from first-principles DFT is described in Secs. II and IV of \cite{Sven_Ga}.

\begin{figure} [h!]
\includegraphics [width=0.35\textwidth]{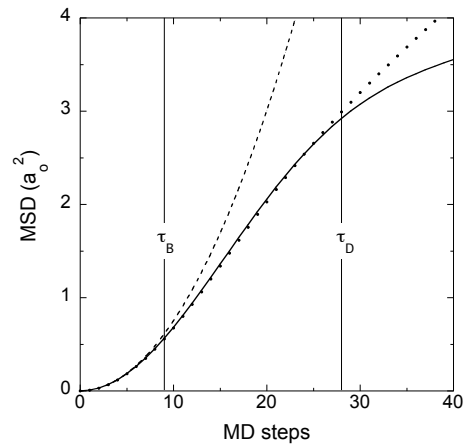}
\caption{Dots are $X_{MD}(t)$, line is  $X_{vib}(t)$ and dashed line is the ballistic contribution, which is accurate only to $\tau_{B}=9~\delta t$.}
\label{fig2}
\end{figure}
Ballistic  motion satisfies ${\bf u}_{K}(t) - {\bf u}_{K}(0) = {\bf v}_{K}(0) t$, where ${\bf v}_{K} $ is the velocity of atom $K$, so the ballistic contribution to the MSD is $(3k_{B}T/M)t^{2}$.  This is precisely the leading term in the small-$t$ expansion of Eq.~(\ref{eq6}): the vibrational motion automatically produces the correct ballistic contribution in $X_{vib}(t)$. As shown in Fig.~\ref{fig2}, the ballistic contribution is accurate for only a small part of the vibrational interval, namely to a time around $\frac{1}{3} \tau_{D}$ and a magnitude around $\frac{1}{5}  X_{vib}(\tau_{D})$. Following the ballistic regime, the terms in $\cos(\omega_{\lambda}t)$ take over and change the curvature of $X_{vib}(t)$ from positive to negative in Fig.~\ref{fig2}. Moreover, the set of $\cos(\omega_{\lambda}t)$ begin to dephase, and this dephasing causes the leveling of $X_{vib}(t)$ until it becomes constant at $\tau_{RW}$ in Fig.~\ref{fig1}.

In Fig.~\ref{fig1}, $X_{vib}(t)$ shows a small slope   discontinuity at $\tau_{RW}$, arising as follows. The actual $X_{vib}(t)$ curve weakly overshoots $X_{vib}(\tau_{RW})$ and rises to a low   maximum around  $150 \delta t$, where $\delta t$ is the MD time step, then falls back to $X_{vib}(\tau_{RW})$ at $300 \delta t$, where it remains. The overshoot is the final dephasing of the lowest frequency vibrational modes. However, we are not able to discern a remnant of this overshoot in $X_{MD}(t)$; the slope of $X_{MD}(t)$ is strictly constant where the vibrational overshoot is present. We attribute this effect to an extra damping of the vibrational motion by the strong transit activity in the MD system at $t \gtrsim \tau_{RW}$. This damping is a vibration transit interaction, which we account for by setting $X_{vib}(t)$  to $X_{vib}(\tau_{RW})$ at $t\geq \tau_{RW}$, causing the slope discontinuity in Fig.~\ref{fig1}.

\begin{table}[ht] 
\caption{\label{table1}Values for the parameters as defined in the text. The MD time step is $\delta t = 7.00288$~fs.}
\begin{ruledtabular}
\begin{tabular}{ccccc}
$\nu$&$\delta R$&$S$&$\tau_{D}$&$\tau_{RW}$\\
\hline
3.9~ps$^{-1}$&1.75~a$_{0}$&1.46~a$_{0}$&$28~\delta t$&$60~\delta t$\\
\end {tabular}
\end{ruledtabular}
\label{Table1}
\end{table}

The question now is, how can the MD system measure nothing but pure vibrational motion for $t$ up to $\tau_{D}$? As illustrated in Eq.~(\ref{eq1}), the MSD measures the motion of one  atom at a time on the system trajectory. This single-atom  motion can be decomposed into continuous vibration about its fixed equilibrium position ${\bf R}_{K}$, plus the mean transit-induced motion of its equilibrium position, abbreviated ``transit motion". With each transit, this mean motion starts from zero and covers a distance denoted $\delta R$. The MSD cannot resolve  this motion, hence cannot measure it, until it has reached a sufficient magnitude. This requires a certain time which we identify as $\tau_{D}$. 

For the collective liquid system, transits are occurring throughout  the volume uniformly in space and time, with mean transit rate $\nu$ per atom. Eventually, $X_{MD}(t)$ is able to fully measure every transit, from start to completion, and the measured collective transit motion becomes a steady process: its effect is to move the equilibrium position of every atom a distance $\delta R$, uniformly distributed over directions, in each time period $\nu ^{-1}$.  This motion is a random walk, whose contribution to the MSD 
starts at $\tau_{RW}$ and thereafter is 
\begin{equation} \label{eq7}
X_{RW}(t) = \nu (\delta R)^{2} (t-t_{RW}),  ~~~~~~~~~~~~t\geq\tau_{RW}.
\end{equation}
This random walk was initially derived as a damping factor for the self intermediate scattering function \cite{us_Fqtself} and was calibrated to  the Einstein relation $\nu (\delta R)^2 = 6D$, which of course is the correct calibration here as well. Moreover, $\delta R$ was also calibrated independently \cite{us_Fqtself}, from our observed transits \cite{Obs_of_tr_2001}, so we have values for both $\nu$ and $\delta R$, which are listed in Table~I.

\begin{figure} [h!]
\includegraphics [width=0.35\textwidth]{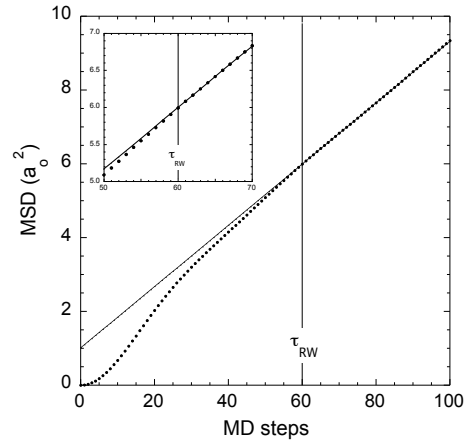}
\caption{Dots are $X_{MD}(t)$ and line is the straight line fitted to $X_{MD}(t)$, shifted to match $X_{MD}(\tau_{RW})$. These curves determine $\tau_{RW}$ to appropriate (qualitative) accuracy by showing that $X_{MD}(t)$ increases with $t$ until it reaches the straight line at $\tau_{RW}$, and then remains on the straight line.}
\label{fig3}
\end{figure}
Finally we can define $\tau_{RW}$ in the context of V-T theory: it is the time when $X_{vib}(t)$ has saturated to a constant and the transit random walk has become a steady process. As a practical way to calibrate $\tau_{RW}$, the slope of $X_{MD}(t)$ should reach $6 D$ at $\tau_{RW}$  and should remain there as $t$ increases. Graphical application of this technique is shown in Fig.~\ref{fig3} and gives the value $\tau_{RW} =   60 \delta t$. 
This evaluation is confirmed below.

\begin{figure} [h!]
\includegraphics [width=0.35\textwidth]{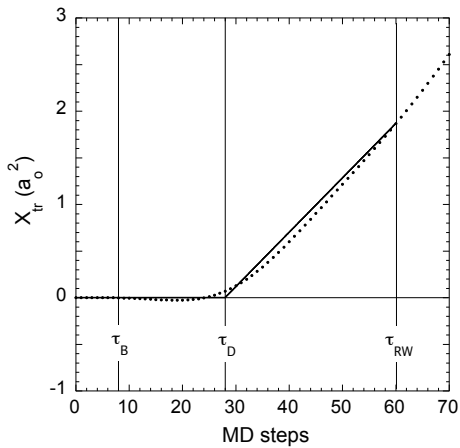}
\caption{Dots are $X_{MD}(t)-X_{vib}(t)$, and the straight line segments show our tractable approximation for $X_{tr}(t)$ up to $\tau_{RW}$.}
\label{fig4}
\end{figure}

To study the transit contribution to the MSD at $t$ up to $\tau_{RW}$, we graph $X_{MD}(t)-X_{vib}(t)$ as the dotted curve in Fig.~\ref{fig4}. The Figure confirms the accuracy of $X_{vib}(t)$ for the ballistic motion. The dotted curve  shows a very small negative dip in the vibrational interval,  the curve then turns sharply upward, but the dominant time dependence is linear in $t$ from $\tau_{D}$ to $\tau_{RW}$.   
All this allows a simple mean-transit approximation for $X_{tr}(t)$ on $0\leq t \leq \tau_{RW}$:
\begin{eqnarray} 
X_{tr}(t) &=& 0,  \qquad    \qquad \qquad \quad0 \leq t \leq \tau_{D};  \label{eq8}\\
X_{tr}(t) &=&\nu (t-\tau_{D})S^{2},\qquad \tau_{D}\leq t \leq \tau_{RW}.\label{eq9}
\end{eqnarray}
In Eq.~(\ref{eq9}), $\nu(t-\tau_{D})$ is the number of transits per atom in $t-\tau_{D}$, and $S^2$ is the mean single-transit contribution over the crossover interval. $S^2$ is calibrated by setting $X_{tr}(\tau_{RW})$ to  $X_{MD}(\tau_{RW})-X_{vib}(\tau_{RW})$, which expresses the endpoint   at $\tau_{RW}$ of the second line segment in Fig.~\ref{fig4}. The error of Eqs.~(\ref{eq8}) and (\ref{eq9}) is seen in Fig.~\ref{fig4} and is assigned to $X_{int}(t)$ and neglected.
We find $S=1.46$~a$_0$, a bit less than $\delta R$ in Table~I because some of the transit contribution is lost in the delay.  
The approximation Eq.~(\ref{eq9}) is specifically in terms of the atomic motion, and is tractable.

To test the stability of the fit provided by Eqs.~(\ref{eq7})~-~(\ref{eq9}), we systematically varied $\tau_{D}$ and $\tau_{RW}$ away from their values in Fig.~\ref{fig4}, and found the overall error increases but remains acceptable for $\tau_{D} = (27-29)\delta t$ and $\tau_{RW}=(56-66)\delta t$. The theory  does not ask for accurate calibration of these characteristic times.

The equations for $X_{VT}(t)$ on the three time intervals of Fig.~\ref{fig1} are summarized as follows.
\begin{tabular}{llr}

1.  Vibrational interval:  \qquad \qquad \qquad$ 0 \leq  t \leq \tau_{D}$     & \\
     $ \qquad  X_{VT}(t)$ =$X_{vib}(t)$& $  ~(9)$\\

2. Crossover interval:  \qquad \qquad  \qquad $\tau_{D} \leq t \leq \tau_{RW} $&\\
     $ \qquad X_{VT}(t)$ =$X_{vib}(t) +  \nu (t-\tau_{D}) S^2$&(10)\\

3. Transit Random Walk interval: \qquad $t \geq \tau_{RW}$&                      \\
    $ \qquad  X_{VT}(t)$ =$X_{VT}(\tau_{RW}) +  \nu (t-\tau_{RW}) (\delta R)^2 $& (11)
\end{tabular}

The following summary of the motion compares directly with the established description cited in the introduction.

 The vibrational motion contributes to the MSD on $t \geq 0$ (Fig.~\ref{fig1}). This motion is ballistic for a very short time, then goes over to vibrational dephasing until its completion at $\tau_{RW}$, when $X_{vib}(t)$ becomes constant. $X_{vib}(t)$ accurately agrees with $X_{MD}(t)$ on $0 \leq t \leq \tau_{D}$, therefore $X_{vib}(t)$ constitutes the complete theory until $\tau_{D}$. 
 
\begin{figure} [h!]
\includegraphics [width=0.35\textwidth]{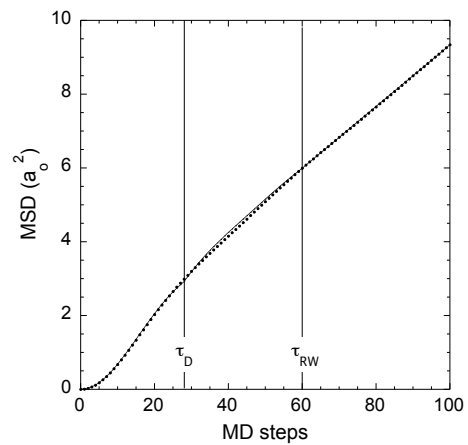}
\caption{Dots are $X_{MD}(t)$ and line is the calibrated $X_{VT}(t)$ theory. Maximum theoretical error is 2.5\% near the center of the  crossover interval, and arises from the neglect of an interaction term.}
\label{fig5}
\end{figure}
Transit motion contributes to the MSD on $t \geq \tau_{D}$ (Fig.~\ref{fig1}), and $X_{tr}(t)$ builds up from zero at $\tau_{D}$ to the ultimate random walk at $\tau_{RW}$ (Fig.~\ref{fig4}). This build up is accurately accounted for: Fig.~\ref{fig5} compares $X_{VT}(t)$ and $X_{MD}(t)$ for all $t$, and shows the maximum error magnitude of 2.5\% near the center of the crossover interval.

At $t \geq \tau_{RW}$, the transit motion constitutes a random walk of the atomic equilibrium positions, which is calibrated from the self-diffusion coefficient $D$, making the theory agree with $X_{MD}(t)$. 

We conclude that this account of the atomic motion is physically sensible because it agrees with the established description, and the account also adds significant  information to that description.

We are pleased to thank Brad Clements, Lee Collins, Joel Kress and Arthur Voter for helpful and encouraging discussions. This research is supported by the Department of Energy under Contract No. DE-AC52-06NA25396.

\bibliography{TheoryforFd} 

\end{document}